# Non-adiabatic suppression of 3D excitonic screening in black phosphorus by mid-infrared pulses

Angela Montanaro[1,2], Francesca Giusti[1,2], Matteo Zanfrognini[3], Paola Di Pietro[2], Filippo Glerean[1,2], Giacomo Jarc[1,2], Enrico Maria Rigoni[1,2], Shahla Y. Mathengattil[1,2], Daniele Varsano[4], Massimo Rontani[4], Andrea Perucchi[2], Elisa Molinari[3,4], and Daniele Fausti[1,2,*]

[1]*Department of Physics, Università degli Studi di Trieste, 34127 Trieste, Italy*
[2]*Elettra Sincrotrone Trieste S.C.p.A., 34127 Basovizza Trieste, Italy*
[3]*Dipartimento FIM, Università degli Studi di Modena e Reggio Emilia, 41125 Modena, Italy*
[4]*Consiglio Nazionale delle Ricerche - Istituto Nanoscienze, 41125 Modena, Italy*

[*]*Correspondence: daniele.fausti@elettra.eu*

**The competition between the electron-hole Coulomb attraction and the three-dimensional dielectric screening dictates the optical properties of layered semiconductors. In low-dimensional materials, the equilibrium dielectric environment can be significantly altered by the ultrafast excitation of photo-carriers, leading to renormalized band gap and exciton binding energies [1,2]. Recently, black phosphorus emerged as a 2D material with strongly layer-dependent electronic properties [3,4]. Here, we resolve the coherent response of screening to sub-gap photo-excitation in bulk black phosphorus and find that mid-infrared pulses tuned across the band gap drive a transient non-thermal suppression of the dielectric screening, which is revealed by the emergence of the single-layer exciton resonance. Our work exposes the role of interlayer interactions in determining the electronic properties of 2D materials and discloses the possibility of optically manipulate them, which is of great relevance for the engineering of versatile van der Waals low-dimensional materials.**

Layered van der Waals (vdW) materials have received increasing attention in recent years due to their potential applications in optoelectronics and solar energy harvesting [5-7]. Due to spatial confinement, the optical response of atomically thin semiconductors is dominated by strongly bound excitons, whose typically large binding energies are drastically reduced by more than an order of magnitude in their bulk counterpart. Understanding how the excitonic structure is affected by the dimensionality crossover from 2D to 3D is of key importance to clarify the role of interlayer coupling and dielectric screening in defining the electronic properties of layered semiconductors. However, this is intrinsically challenging in most of 2D materials, like transition-metal dichalcogenides (TMDCs) and hexagonal boron nitride (h-BN) semiconductors, mainly because the band gap is direct only in the monolayer, while indirect in the bulk material. Black phosphorus (BP) has recently emerged as a promising candidate to bridge this gap [3]. With a carrier mobility comparable to that of graphene, BP features a direct band gap in both its single-layer (phosphorene) and bulk form [3,4]. The amplitude of the gap is strongly layer-dependent, spanning from the visible (2 eV) to the mid-infrared (0.3 eV) range as the layer thickness is increased, thereby making BP a unique platform to study the dimensionality crossover from 2D to 3D.

The application of external stimuli, such as electric fields [8], pressure [9,10], in-plane mechanical strain [11-14] and dopants [15,16], has proved an effective way to manipulate the electronic structure of BP. Ultrashort pulses are emerging as an exceptional tool to address the non-equilibrium response and possibly control both its electronic and optical properties on the ultrafast time-scale [17-22]. In particular, photo-excitation by near-infrared laser pulses has been found to trigger a band gap renormalization in bulk BP due to the transient enhancement of dielectric screening induced by the excited photo-carrier population [23-25]. However, the non-equilibrium dielectric environment upon photo-injection of a small (i.e., smaller than the band gap) excess energy, remains unexplored.

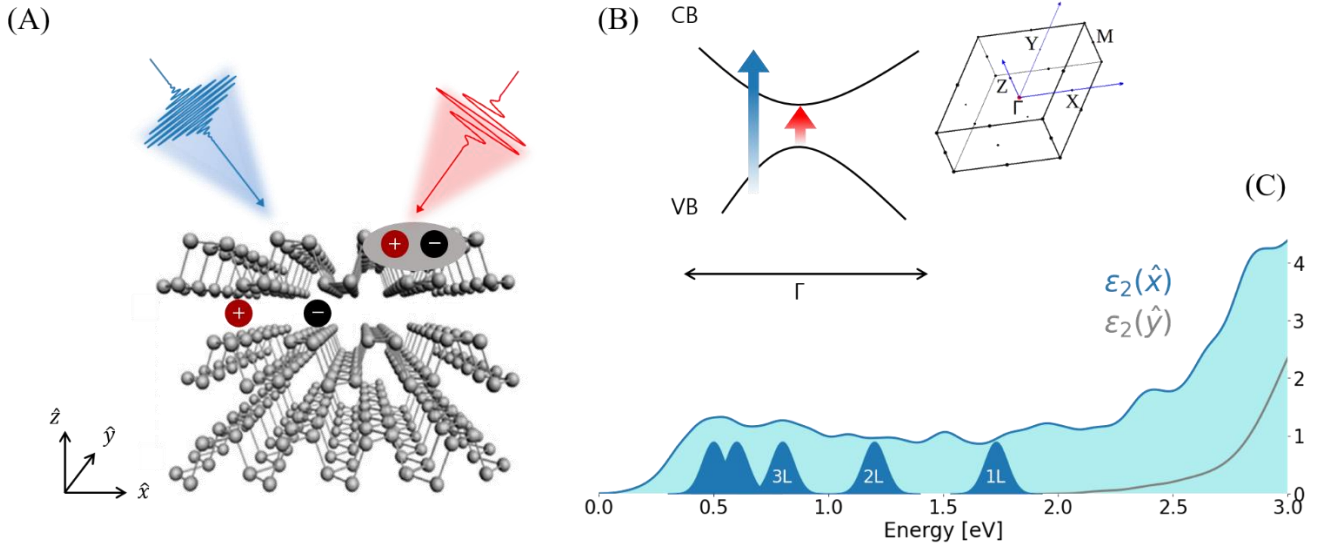

**Fig. 1: Optical fingerprint of 3D screening in bulk black phosphorus.** (**A**) Cartoon of the transient screening suppression by mid-infrared pulses revealing the undressed exciton in the monolayer phosphorene. BP crystal structure was adapted from ref. [26]. (**B**) Simplified sketch of the electronic structure at the Γ point, as indicated in the first Brillouin zone. The arrows represent the high-photon energy (blue) and sub-gap (red) photo-excitation. (**C**) Calculated imaginary part of the dielectric function in bulk BP along the armchair ($\hat{x}$, blue) and zigzag ($\hat{y}$, grey) direction, convoluted with a numerical broadening of 100 meV. The blue peaks denote the lowest-energy exciton resonances ($E_{11}$) in monolayer (1L), bilayer (2L), trilayer (3L) etc. BP, adapted from ref. [32].

In this work, we investigate the non-equilibrium response of dielectric screening in bulk BP to both above-gap and sub-gap photo-excitation. Experimentally, we photo-excite the sample by ultrashort pulses with photon energy tunable across the equilibrium mid-infrared (MIR) band gap and monitor the photo-induced change in reflectivity over a broad energy range (Fig. 1A,B). We find that photo-excitation by high- and low-photon energy pulses yields remarkably different optical responses. While high-photon energy excitation leads to a broadband light-induced transparency in the visible range due to phase space filling, excitation with photon energies comparable to the band gap triggers an effective suppression of screening, which is revealed by the appearance of the undressed single-layer exciton dynamics. With the support of first-principle DFT calculations, our measurements suggest that the observed reduced screening environment may be triggered by MIR-driven ac-currents through the coherent redistribution of the charge carriers. These findings unlock a new route for the optical manipulation of 2D materials and exciton-based applications to optoelectronic switches.

BP is an elemental semiconductor which crystallizes in a layered orthorhombic structure, where, because of sp3 orbital hybridization, phosphorus atoms are arranged in a puckered honeycomb lattice [27]. As shown in Fig. 1A, the resulting layers have two inequivalent high-symmetry directions, the so-called armchair ($\hat{x}$) and zigzag ($\hat{y}$) directions. This strong in-plane anisotropy has macroscopic consequences on both the electronic and optical properties of BP [4,28,29]. We show in Fig. 1C the single-particle optical absorption in bulk BP calculated through first-principle hybrid-functional DFT theory (see Methods). As a result of the symmetry selection rules, light polarized along the zig-zag direction is expected to be absorbed only above ∼2 eV. Conversely, when the polarization is parallel to the armchair direction, the computed absorption threshold is 0.345 eV. This agrees with previous experimental and theoretical studies [4,30], and corresponds to a direct electronic band gap at the Γ point (Fig. 1B)[1].

---

[1] It should be noted that we used in the calculation a conventional unit cell with double sized interlayer distance, as in ref. [31].

In order to identify the optical transitions that give rise to the structured absorption in Fig. 1C, it is crucial to consider the evolution of the optical properties of BP as function of the sample thickness. The direct band gap energy in the monolayer phosphorene is ~2 eV, and the optical absorption is dominated by excitons with binding energies as large as hundreds of meV [32-35]. Although distinct higher-order transitions are theoretically predicted in the single-layer limit [4], the most prominent resonance is the lowest-energy exciton ($E_{11}$), that has been identified by photoluminescence and absorption spectroscopy, and lies at ~1.73 eV [32,36]. Importantly, as the layer number (L) is increased, the $E_{11}$ resonance (along with the band gap) monotonically shifts to lower energies, as a consequence of the enhanced screening of the electron-hole Coulomb attraction. As denoted by the position of the blue peaks in Fig. 1C, the singlet exciton red-shift does not scale linearly with the layer number and eventually reaches a plateau at ~15L, giving rise to the absorption edge in the bulk limit (0.34 eV) [33]. This critical thickness is consistent with the theoretically predicted screening length of 10 nm, which is roughly 20 atomic layers [4]. The optical absorption of bulk BP is thus the result of the strong interlayer interactions that arise from the reduced perpendicular quantum confinement. In particular, the optical transitions within 0.34 eV (gap energy in bulk BP) and 2 eV (gap energy in the monolayer) are intrinsically related to the three-dimensional nature of the material [37]. This is evident in our DFT calculations, which confirm that absorption in this optical range is dominated by transitions involving solely the lowest energy bands dispersing along the stacking direction, as detailed in the Supplementary Materials (SM, Note 11). In our time-domain measurements, we leverage on this characteristic and use a white-light supercontinuum (1.3-2.2 eV) to monitor the transient reflectivity in this spectral region that is expected to be the most sensitive to pump-induced changes of the 3D screening environment.

The broadband transient reflectivity measurements were carried out on freshly cleaved bulk BP employing the experimental setup described in ref. [38]. From the steady-state Fourier transform infrared (FTIR) measurements reported in Fig. S1, we identified the gap energy at room temperature to be ~0.33 eV. In contrast to most of semiconductors, the band gap in bulk BP redshifts at lower temperatures, to reach ~0.28 eV at 12 K. This anomalous behavior has been widely observed [39-42], and only recently related to the temperature-dependence of the interlayer vdW coupling [43]. We used the FTIR measurements as a benchmark to tune the photon energy of the pulses that drive the sample out of the equilibrium conditions. We investigated two distinct regimes: a) the photo-injected excess energy is larger than the band gap ($h\nu > E_g$), so a carrier population is excited in the material; b) the sample is photo-excited by MIR pulses ($h\nu < E_g$) that are not energetic enough to initiate electronic transitions.

We show in Fig. 2C the time- and energy-resolved transient reflectivity change upon high photon-energy excitation. The measurement reported was performed at T=10 K, but we stress that this result is not affected by the sample temperature (Fig. S5). For probe photon energy ($E_{pr}$) below 2 eV, the pump-probe time traces are characterized by an initial negative change in reflectivity, followed – after approximately 1 ps – by a less intense positive signal (Fig. 2D), that lasts for hundreds of ps (Fig. S2). These results are consistent with previous quasi-monochromatic pump-probe experiments in the near-infrared range, in which the early-time signal is ascribed to photo-bleaching due to Pauli blocking and the subsequent one to photo-induced absorption by the excited free carrier population [17,18,20].

Our broadband measurements, however, put these interpretations in a new perspective and show that the early-time photo-induced transparency is not a universal feature of the ultrafast response of bulk BP, but that it vanishes for $E_{pr} > 2$ eV. This is an indication that the Pauli blockade effect induced by above-gap photo-excitations targets only the lowest energy optical band associated to the dispersion along the stacking direction. To clarify this point, we compare in Fig. 2A the DFT optical absorption computed including in the calculation all the optical transitions (black line), and only the transitions between the last valence band (lVB) and the first conduction band (fCB) (yellow-shadowed area). The two curves overlap up to approximately ~1.7 eV and then start to deviate, until the contribution of the optical transitions from lVB to fCB is almost suppressed. As transitions to other bands dominate above 2 eV, the probe can be absorbed even if the lVB and fCB are already occupied by the photo-excited carrier population, thereby overcoming the Pauli blocking effect. We stress that this effect is not related to a specific pump photon energy and photo-excitation by pumps with smaller photon energy (0.65 eV) but still larger than the band gap, induces a similar spectral response (Fig. S4).

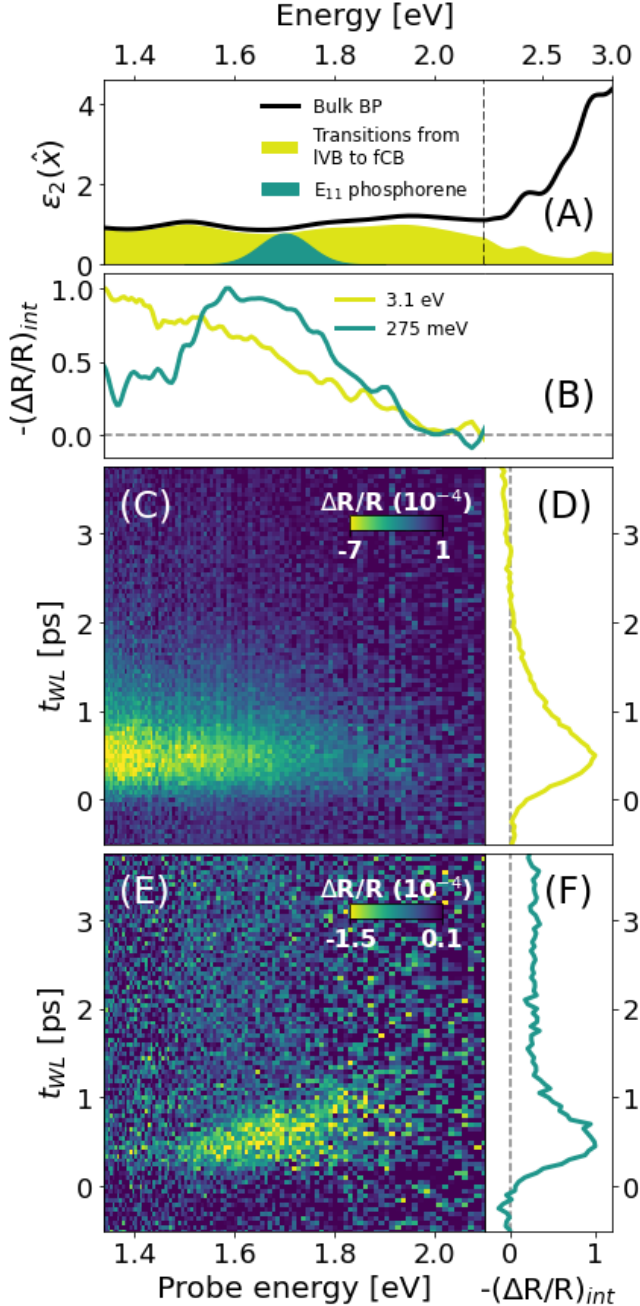

**Fig. 2: Above- and below-gap photo-excitation in bulk black phosphorus.** (A) DFT-calculated total optical absorption of bulk BP (thick black line). The yellow area indicates the optical absorption obtained when only the transitions from the last valence band (lVB) to the first conduction band (fCB) are included in the calculation. The green Gaussian shape indicates the lowest-energy exciton resonance in the single-layer limit. The energy scale is uniformly spaced from 1.34 eV up to 2.1 eV (vertical dashed black line), and it is shrinked above up to 3 eV. (C),(E) Transient reflectivity map measured on bulk BP at 10 K upon photo-excitation by high-photon energy (3.1 eV) and sub-gap (275 meV) pulses, respectively. The pumping fluences were 21 µJ cm$^{-2}$ and 130 µJ cm$^{-2}$. (B) Normalized spectra at fixed $t_{WL}$=500 fs of the maps in (C) and (E). A Gaussian smoothing has been applied to both traces. (D),(F) Normalized pump-probe traces of the maps in (C) and (E) integrated over the region (1.4-1.7 eV) and (1.5-1.8 eV), respectively.

The broadband photo-induced transparency below 2 eV is not present when the sample is excited by pulses with photon energy smaller than the IR gap (Fig. 2E). Strikingly, the spectral response with long wavelength pumps displays a qualitatively different response, that is characterized by a negative feature peaked at ~1.7 eV (Fig. 2B) and exhibits a long-lived dynamics (Fig. S3). The suppression of the Pauli blocking effect is an indication that no optical transitions occur along the stacking direction upon the photo-excitation. Moreover, the emergence of a transient response that is well localized in frequency is in stark contrast with the calculated optical absorption of bulk BP (Fig. 2A), that does not display any particular resonance at 1.7 eV. Based on these arguments and the assignments in refs. [32,36,44], we identify this signal with the appearance of a transient absorption from the lowest-energy exciton resonance in the monolayer phosphorene that indicate a suppression of the dielectric screening. In contrast to previous transient absorption measurements in few-layer BP that reported a derivative-like signal at the exciton resonance [20], our measurements display a signal centered at the expected $E_{11}$ transition, with no significant energy shifts within the time window considered in Fig. 2E. This means that, at least at early-times, many-body interactions (such as renormalization of the band gap and exciton binding energy, which would determine a pump-induced energy shift in the data) are negligible. A blue-shift of the resonance is found for $t_{wL} >$ 15 ps (Fig. S3) and attributed to carrier-phonon interactions, as already observed in other layered semiconductors [45,46].

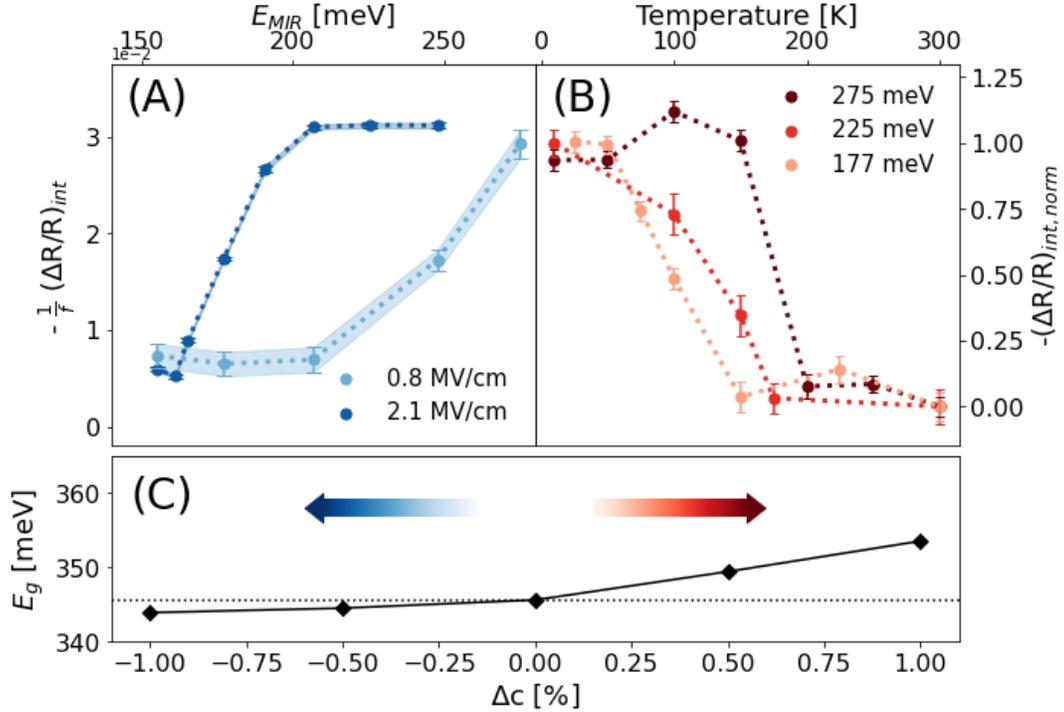

**Fig. 3: Non-adiabatic control of the black phosphorus gap with mid-infrared pulses.** (A) Intensity of the exciton resonance at 10 K as function of the photon energy of the MIR pump in a low (130 µJ cm$^{-2}$, light blue line) and high (890 µJ cm$^{-2}$, dark blue line) field strength regime. Each point in the plot is the result of an integration over a specific probe energy- and time-range (1.4-1.8 eV and 0-750 fs, respectively). The curves are divided by the corresponding fluence. (B) Normalized intensity of the exciton resonance as function of the sample temperature for three different photon energies of the MIR pump. (C) DFT-calculated band gap energy in bulk BP with modified interlayer distance (c). Negative (positive) values of Δc correspond to a compression (expansion) along the stacking direction.

In order to shed light on the mechanism leading to the observed screening suppression, we studied the amplitude of this anomalous response at different pump photon energy across the band gap of bulk BP and quantified the intensity of the transient signal associated to the exciton resonance (as detailed in SM, Note 9). We plot in Fig. 3A the results of this analysis for two different fluence regimes of the MIR pump pulse. We observe that the exciton signal is quenched at small pump photon energies. In particular, the exciton is suppressed at approximately 200 meV at low MIR fluences, while the cut-off edge is redshifted at higher pumping fluences. We associate this cut-off to the effective band gap energy of the material upon the MIR photo-excitation, which is found to trigger a transient gap closure. While it would be tempting to ascribe the redshift of the response to a carrier-induced band gap renormalization, as observed in ref. [24], we stress that the photo-injected excess energy here is smaller than the equilibrium band gap and no above-gap free-carrier population is excited in the linear response. In order to rule out population effects due to two-photon absorption, we studied the pump fluence-dependence of the response to MIR excitation. As evidenced in Fig. S7, the signal associated to the exciton resonance does not scale as the square of the fluence (as it would be expected if two-photon absorption processes were involved), but it is consistent with a coherent effect that scales with the amplitude of the electric field.

While the further redshift of the band gap at higher MIR fluence (Fig. 3A) could suggest a thermally driven effect, we ruled out this possibility by carrying out temperature-dependent measurements at fixed MIR pump photon energy. In Fig. 3B, each point indicates the intensity of the exciton resonance measured at a given temperature and pump photon energy (details in SM, Note 10). For all the three examined MIR photon energies, the exciton disappears at high temperature, hence providing a clear indication that its emergence is related to the equilibrium band gap energy. Importantly, the high temperature cut-off for the dynamical exciton signal occurs at higher temperature when the MIR photon energy is larger. The temperature-dependence of the gap energy is then consistent with the equilibrium thermoelectric properties of BP, but follows an opposite trend with respect to the MIR fluence-dependent measurements in Fig.3A.

Our findings suggest that the transient suppression of screening driven by MIR excitation is accompanied by a non-thermal closure of the infrared band gap. To identify a possible mechanism that could explain our experimental evidence, we ran a set of simulations with modified interlayer spacing ($c$) in bulk BP at equilibrium. The rationale of this approach is that the 3D screening (along with the corresponding band structure and gap energy) at equilibrium is ultimately set by the interlayer distance in the simulated material. By freezing the in-plane atomic distances, we calculated the band gap energy when the crystalline structure is compressed ($\Delta c<0$) or strained ($\Delta c>0$) along the stacking direction up to 1% (Fig. 3C). We found that when the interlayer distance is increased, the gap energy increases as well, in agreement with previous calculations [4]. This scenario is qualitatively analogous to an adiabatic thermal expansion, which, as observed both at equilibrium (Fig. S1) and in temperature-dependent pump-probe measurements (Fig. 3B), is related to an enlargement of the infrared optical gap. The observed MIR-driven collapse of the band gap (Fig. 3A) is instead compatible with a contraction along the stacking direction.

We stress, however, that such comparison is only qualitative, while quantitatively the observed dynamical collapse of the infrared gap remains unexplained and cannot be rationalized solely as an adiabatic compression of the lattice. On the one hand, the calculated gap energy is much larger than the one extracted by the pump wavelength-dependent experiments. In this regard, it should be noted that in the simulated adiabatic compression the in-plane lattice parameters have been kept fixed, while further in-plane contractions could result in a larger band gap closure [47], as detailed in SM Note 12 where hydrodynamic pressure-dependent simulations are discussed. On the other hand, the most relevant aspect distinguishing the observed MIR-driven response from an adiabatic contraction of the c-axis is the fact that the latter is expected to increase the screening of excitonic resonance, while the optical measurements revealed a dynamical reduction of screening.

Based on the evidence collected so far, we speculate that the sudden suppression of the 3D screening that unveils the exciton in our measurements could be related to a quasi-adiabatic coherent redistribution of the charge carriers driven by MIR pulses. In a simple picture, the bulk material at equilibrium can be seen as an electron bath in which the single-layer excitons are largely screened due to the high carrier mobility. Upon photo-excitation by MIR pulses, the material interacts with a long-wavelength in-plane electric field, which forces the electrons to a coherent motion, resulting in light-driven ac-currents. Due to the sudden charge redistribution, the electron bath is then less effective in screening the exciton, whose transient absorption is then detected by the broadband probe. Moreover, the establishment of such currents might also affect the layered structure of the material. Analogously to the magnetic attraction experienced by wires carrying homodirectional dc-current (Biot-Savart law), the driven in-plane transport of charge carriers may result in a contraction along the stacking direction, that, based on our DFT simulations, could justify the observed band gap closure. While this is a tentative explanation, investigating the coherent response triggered by intense THz fields might be decisive to assess the role of light-driven ac-currents in a layered semiconductor prototype such as BP.

In summary, we have studied the non-equilibrium dielectric environment in bulk BP upon photo-excitation by energy-tunable ultrashort pulses. Our *ab-initio* DFT calculations have shown that the optical absorption below 2 eV is dominated by transitions involving energy bands dispersing along the stacking direction, making this spectral region intrinsically linked to the enhanced 3D screening in the bulk material. We have found that high-photon energy excitations uniformly target this whole band, leading to a spectrally flat photo-induced transparency through Pauli blocking. This bulk-like spectral response is suppressed when the sample is photo-excited by sub-gap mid-infrared pulses, which reveal instead a transient response peaked at the lowest-energy exciton resonance in the monolayer phosphorene. By unveiling how low energy ac-currents reduce the screening of excitonic resonances in quantum materials, our findings disclose the possibility to obtain an ultrafast control of screening in layered semiconductors, which is of paramount importance for 2D materials-based optoelectronic applications.

**METHODS**

PUMP-PROBE EXPERIMENTS
We performed broadband non-equilibrium reflectivity measurements on bulk BP at different temperatures (10 K-300 K). The experimental apparatus was pumped by a Pharos Laser (by Light Conversion), which delivers 400 µJ pulses with 1.2 eV photon energy. We photo-excited the sample by tunable, carrier-envelope phase stable mid-infrared pulses (155-275 meV) obtained by difference frequency generation in a GaSe crystal of two near-infrared pulses, both generated by a Twin Optical Parametric Amplifier (Orpheus TWIN by Light Conversion). The wavelength of the mid-infrared pulses was measured by a home-built Michelson interferometer, equipped with a mercury cadmium telluride (MCT) detector. The 3.1 eV-pump was obtained by second harmonic generation in a β-barium borate crystal of the output of a Non-Collinear Parametric Amplifier (Orpheus-N by Light Conversion). Out-of-equilibrium reflectivity was probed by a supercontinuum white-light extending from 1.3 eV to 2.2 eV generated by self-phase modulation of the Pharos output in a sapphire crystal. The reflected probe beam was diffracted, and its frequency components were acquired by a linear array of silicon photodiodes (NMOS by Hamamatsu) synchronized with the laser repetition rate. The signal-to-noise ratio was increased by the simultaneous acquisition and division by a reference beam (not interacting with the sample) to suppress the white-light fluctuations. For a full description of the experimental apparatus, see ref. [38]. The laser repetition rate was set to 5 kHz. We report that higher repetition rates induce the onset of long-lasting thermal signals that obscure the fast dynamics. The probe was polarized along the armchair direction ($\hat{x}$), while both the visible and the mid-infrared pumps were cross-polarized. The sample axes were identified by polarization-dependent visible pump-probe measurements, according to ref. [18]. The FWHM of the probe beam was 30 µm, small enough to ensure a uniform illumination of the irregular sample surface. No physical correction of the temporal chirp of the broadband white-light probe was performed, but all the data were post-processed to compensate for the dispersion. Samples were provided by HQ graphene and glued on an oxygen-free copper substrate. All the experiments were performed on freshly cleaved multi-layer BP by *in situ* mechanical exfoliation. Samples were kept under vacuum (<$10^{-8}$ mbar) for the entire duration of the experiment to prevent air degradation and cooled down to cryogenic temperatures using a closed-cycle liquid helium cryostat.

*AB-INITIO* CALCULATIONS
The ground state structural and electronic properties of bulk BP have been evaluated at the DFT level, using the Quantum Espresso package [48].
The structural relaxation has been performed using the PBE [49] approximation for the exchange correlation potential together with the plane wave basis set, while ONCVPSP [50] norm-conserving pseudopotentials have been adopted to model the electron–ion interaction; the kinetic energy cutoff for the wave-functions has been fixed to 90 Ry, while the Brillouin zone has been sampled with a 16x16x8 k-point grid. Van der Waals interactions between layers has been included using the Grimme-D2 parametrization [51].

Both atomic positions and lattice parameters (a conventional orthogonal unit cell has been used, with eight atoms per cell) were relaxed up to when forces acting on each atom were below $3 \times 10^{-4}$ eV/Å. The obtained lattice parameters a = 4.43 Å, b = 3.33 Å, c = 10.48 Å are in good agreement with experimental [52] and theoretical results [53].

Hybrid-DFT (using Gau-PBE hybrid functional [54]) has been used to perform electronic band structure calculations. The equilibrium charge density and electronic Kohn-Sham states have been computed using a 20x20x10 k-point grid to sample the Brillouin zone, in combination with a 4x4x2 grid for the sampling of the Fock operator. The obtained direct electronic band gap of 0.34 eV (located at the Γ point) turns out to be in reasonable agreement with the experimental one [55].

Optical properties have been evaluated using the Yambo code [56], at the independent-particle level. The dielectric function has been calculated as

$$\varepsilon_\alpha(E) = 1 + \frac{16\pi}{\Omega} \sum_{c,v} \sum_k \frac{1}{E_{c\mathbf{k}} - E_{v\mathbf{k}}} \frac{|r_{vc}^\alpha|^2}{(E_{c\mathbf{k}} - E_{v\mathbf{k}})^2 - (E + i\gamma)^2}$$

where $E_{c\mathbf{k}}$ ($E_{v\mathbf{k}}$) are the conduction (valence) electronic states computed at the Gau-PBE level, $r_{vc}^\alpha$ are the interband electric dipoles between a valence and a conduction state projected along direction $\alpha$ (the direction of light polarization) and computed using Covariant approach [56]; finally, $\gamma$ is a broadening, here always fixed to 0.1 eV.

## ACKNOWLEDGEMENTS
This work was mainly supported by the MIUR through the PRIN program No. 2017BZPKSZ. D.F. gratefully acknowledges funding from the European Commission through the projects INCEPT (ERC-2015-STG, Grant No. 677488) and COBRAS (ERC-2019-PoC, Grant No. 860365).

## SUPPLEMENTAL MATERIALS

1. **Static characterization: temperature-dependent FTIR measurements**

We characterized the sample by performing steady-state reflectivity measurements in the far- and mid-infrared spectral range at different temperatures. The measurements were performed at the SISSI infrared beamline [1] of the Elettra synchrotron in Trieste (Italy). Reflectivity data were collected using a Bruker Vertex 70v interferometer. The reflectivity is rather flat above 4000 cm$^{-1}$ and reaches a value close to 0.3 for all the temperatures examined (Fig. S1A). In the mid-infrared region, from 500 to 4000 cm$^{-1}$, we observe an increase in the reflectivity which features a mild dependence on the sample temperature. In the far-infrared range, the reflectivity is dominated by a peak localized at ∼130 cm$^{-1}$, that we assign to the $B_{1u}$ IR-active optical mode, in agreement with previous studies [2-5].

From the reflectivity data, we retrieved the optical conductivity for the whole set of temperatures through the Kramers-Kronig relations, as shown in Fig. S1B. On top of the phonon mode (more prominent at low temperatures), we reveal a Drude-like contribution in the low-frequency range, due to the presence of free charge carriers. At higher frequency, a significant rise in the conductivity marks the gap energy of the sample at approximately 2250 cm$^{-1}$ (∼280 meV). The edge clearly shifts towards higher frequencies as temperature increases, confirming the anomalous temperature-dependence of the band gap that has been widely reported [6-10]. Our results are in good agreement with previous optical studies (see ref. [11] and references therein, and refs. [8,12]), although we observe a less sharp step-like edge, which is probably due to the use of unpolarized light in our measurements.

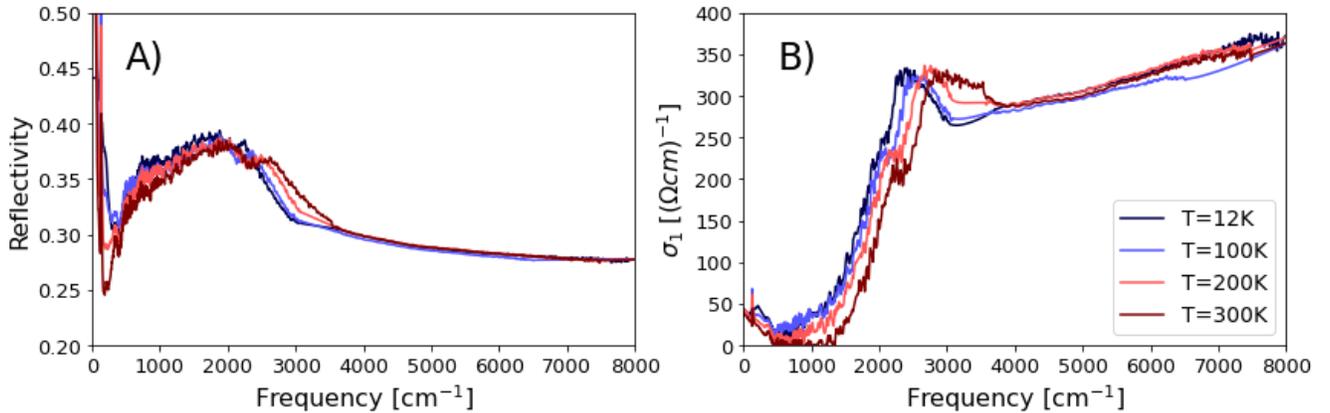

**Fig. S1: FTIR static measurements. A)** Steady-state reflectivity measurements for a variety of temperatures in the mid- and far-infrared spectral range. **B)** Temperature-dependence of the optical conductivity extracted from d) through the Kramers-Kronig relations.

2. **Coherent longitudinal acoustic phonons generation in BP**

Fig. S2A shows the time- and spectrally-resolved relative reflectivity induced by a 3.1 eV photo-excitation for pump-probe delays up to 100 ps. After the initial photo-bleaching discussed in the main text, at longer time scales, the reflectivity is modulated by oscillations, whose frequency increases with increasing probe photon energies. This is better highlighted in Fig. S2B where we plot pump-probe traces for selected probe photon energies averaged over 70 meV. We attribute this energy-dependent modulation of the reflectivity to the onset of Coherent Longitudinal Acoustic Phonons (CLAP) in bulk BP. CLAP detection via pump-probe spectroscopy is a largely studied phenomenon and a well-established tool to measure the longitudinal sound velocity in crystals [13-16]. In this framework, the pump pulse initiates a travelling strain wave which propagates away from the surface at the longitudinal sound velocity of the material and periodically shapes its dielectric function. When the probe impinges on the sample, reflection from both the surface and the CLAP oscillations will contribute to the measured reflectivity, resulting in interferential processes that cause the oscillatory behaviour observed in Fig. S2B. Being the result of an interferential process, the oscillating frequency ($f$) depends on the wavelength ($\lambda$) of the probe pulses, according to the following relation [14]:

$$f = \frac{2nv_s}{\lambda}$$

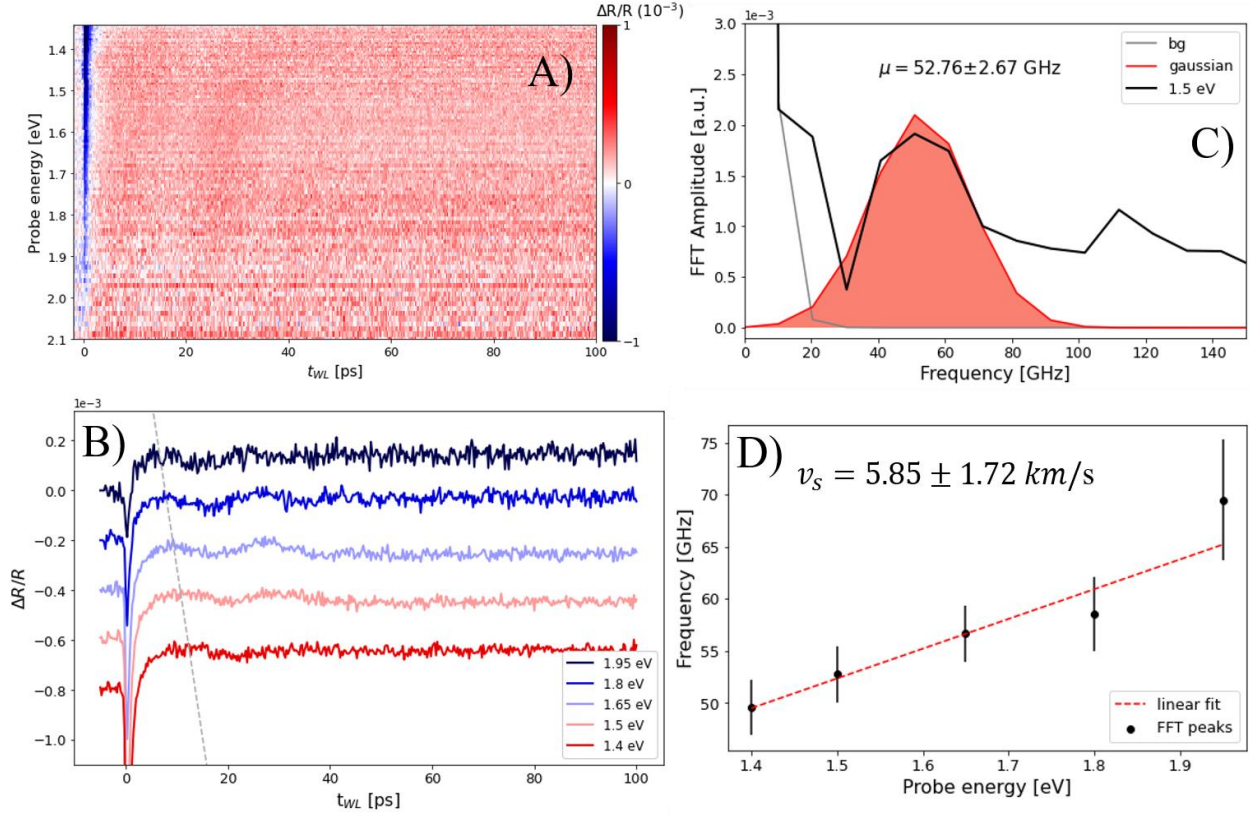

**Fig. S2: Longitudinal sound velocity revealed by Coherent Longitudinal Acoustic Phonon (CLAP) creation. A)** Transient change in reflectivity induced by visible ultrashort pulses as function of time-delay and spectral content of the white-light supercontinuum probe. **B)** Horizontal cuts of the map in A) for a selection of probe photon energies, each averaged over 70 meV (±35 meV with respect to the value indicated in the label). The traces are arbitrarily shifted for clarity. The dashed grey line is a guide for the eye to highlight the linear dependence of the CLAP frequency on the photon energy of the probe. **C)** Fourier transform (black line) of the 1.5 eV curve in B). The Fourier transform calculation has been performed avoiding the early-time negative dynamics of the signal. The peak arising at about 50 GHz was fitted by a gaussian-like function (red shape) to estimate its central frequency. **D)** Central frequency of the Fourier-transform estimated as in C) as function of the selected probe photon energies (black bullets). The dashed red line indicates a linear fit to the data, whose slope gives an estimation of the longitudinal sound velocity of black phosphorus through Equation S1 [14].

(Eq. S1)

where $n$ is the refractive index and $v_s$ the longitudinal velocity.

The technique has been applied recently to bulk BP to study how the in-plane anisotropy affects the CLAP generation [17]. Here, we will use the wavelength-dependent reflectivity oscillations to estimate the sound velocity and provide a characterization of the sample.

For this purpose, we estimate the oscillating frequency by Fourier-transforming the time-traces in Fig. S2B. We discard in the calculation the initial transient response. An example of the Fourier analysis is given in Fig. S2C, where the black line is the Fourier-transform of the oscillating reflectivity measured at $h\nu$=1.5 eV. We fit the peak arising at about 50 GHz with the sum of a gaussian function (red shape) and an exponential decay to account for the incoherent contributions. By repeating the same procedure for all the time-traces in Fig. S2B, we get the oscillating frequencies that we plot in Fig. S2D as function of the probe photon energy. A linear fit to the data estimates the longitudinal sound velocity in our sample to be $v_s$ = 5.85±1.72 km/s, in agreement with the literature [17,18]. In the calculation, we considered the refractive index calculated by ref. [19].

### 3. Transient optical response to sub-gap photo-excitation for long delay time

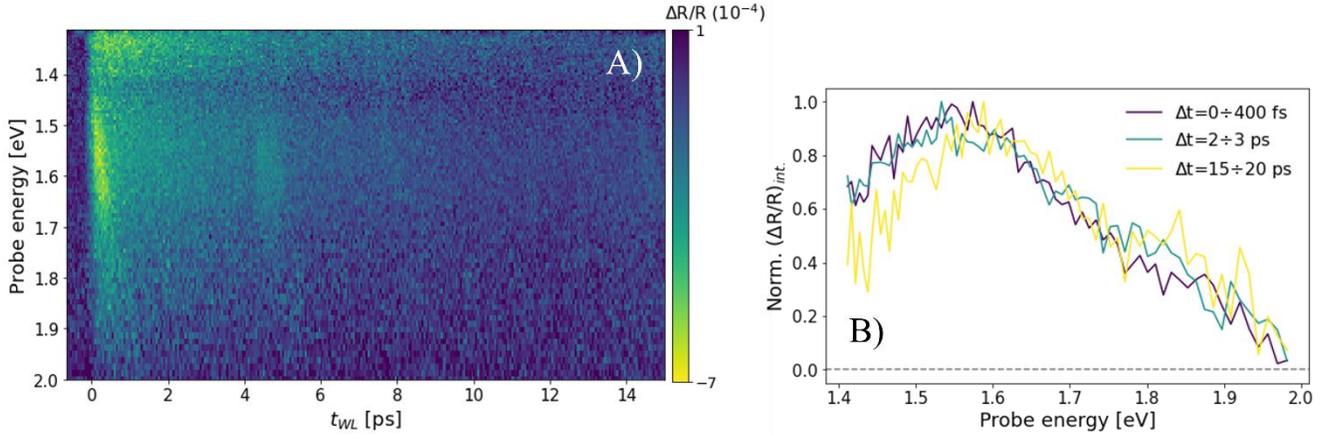

**Fig. S3: Long delay time dynamics.** (**A**) Broadband transient reflectivity measured on BP at T=10 K at long delay time following a mid-infrared photo-excitation (225 meV). The pump fluence was 0.89 mJ cm$^{-2}$. The signal arising at $t_{wL}\simeq$ 5 ps is due to a residual reflection coming from the copper substrate and will therefore be neglected in the discussion. (**B**) Normalized ΔR/R spectra of the excitonic resonance averaged at different time delays, as denoted by the legend. The two traces at early times are centered at the same energy, while the trace at longer delay time is blue-shifted by ∼20 meV.

We show in Fig.S3A the time- and frequency-dependent reflectivity measured at T=10 K up to 15 ps upon photo-excitation by sub-gap mid-infrared pulses (225 meV). The fluence of the mid-infrared pump was higher than that used in Fig.2B in the main text, and accounts for a more intense background signal on top of the excitonic resonance at ∼1.6 eV. A less prominent replica of the exciton is also visible at $t_{wL}\simeq$ 5 ps and is due to a small fraction of the pump that is transmitted through the sample and reflected back by the substrate.
The temporal evolution of the optical response is the result of different many-body processes that modify both the spectral shape and the central energy of the excitonic resonance. In particular, exciton-exciton interactions and free carrier-induced band gap renormalization dominate at early times, and may result in both a transient modification in the oscillator strength and an energy shift of the exciton resonance [20]. At longer delay times (>10 ps), the interplay between the photo-excited carriers and the phonons becomes dominant. The excess energy is transferred to the lattice, which undergoes a transient temperature increase. This results in a non-equilibrium renormalization of the exciton resonance which is shifted by few meV [21].
In Fig.S3B we show the frequency-resolved normalized spectra of the exciton at selected time delay. While the response is identical at early times (purple and green traces), the resonance is blue-shifted by ∼20 meV at $t_{wL}$> 15 ps (yellow trace). This is consistent with the long-time dynamics of the excitonic resonances measured on other layered semiconductors, such as WSe$_2$ [20] and MoS$_2$ [22], and reinforces our assignment of the sub-gap pump-induced excitonic resonance.

## 4. Above-gap photo-excitation by near-infrared pulses

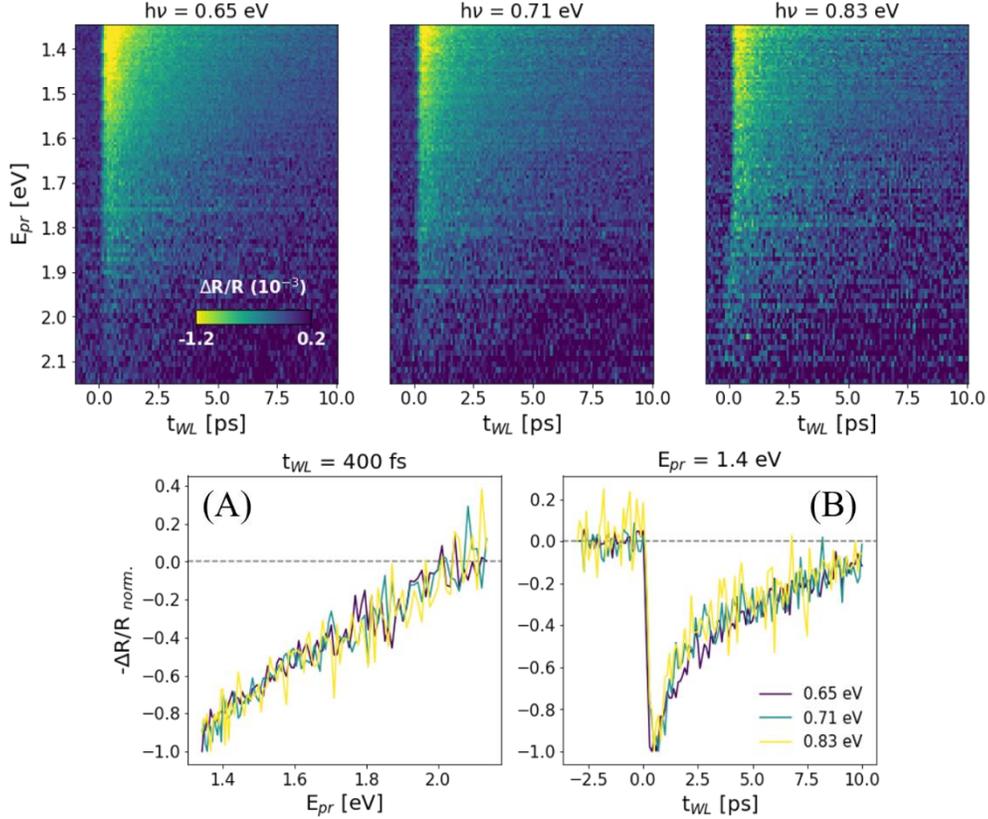

**Fig. S4: Photo-excitation by near-infrared ultrashort pulses.** Transient reflectivity color-coded maps upon photo-excitation by near-infrared pulses with tunable photon energy. **(A)** Normalized probe energy-dependent spectra at fixed time delay ($t_{WL}$=400 fs) for different pump photon energy. **(B)** Normalized pump-probe traces at fixed probe energy ($E_{pr}$=1.4 eV).

In order to explore the non-equilibrium optical response upon injection of different amounts of excess energy, we photo-excited the sample with near-infrared ultrashort pulses. Considering the characteristic mid-infrared band gap of bulk BP, these photon energies are large enough to excite a photo-carrier population. At the same time, they are much lower than the visible photo-excitation discussed in the main text that may eventually initiate higher-order electronic transitions.

We used one of the two near-infrared outputs of the Twin Optical Parametric Amplifier as pump and tuned it in the range 0.65-0.83 eV. We show in Fig. S4 the results of the experiment. There is no appreciable difference neither in the spectral shape nor in the dynamics of the signal upon photo-excitation by different photon energies (Fig. S4 A,B). The spectral dependence of the transient reflectivity is very similar to the one photo-excited by visible pulses: at early times, there is a broadband photo-bleaching (negative differential reflectivity). Similar to the optical response to 3.1 eV photo-excitation discussed in the main text, the photo-bleaching is confined below ~2 eV. Above this threshold, the optical absorption is dominated by higher-order transitions, which are not affected by Pauli blocking. However, when compared to Fig. 2A, the dynamics of the early-time photo-induced transparency due to phase space filling is slower at smaller pump photon energy. This difference could be explained as follows. The photo-bleaching by Pauli blocking is ultimately due to the fact that the probing energy levels are already occupied by the pump-excited carriers. On a picosecond time scale, intra-band scattering with phonons leads to a relaxation of the free photo-carriers that results in a reduction of the Pauli blocking contribution to the optical signal. As the pump photon energy is decreased, also the number of de-excitation channels available for the free carrier population is reduced. This could result in a slower electron-phonon scattering dynamics and, in turn, in the observed slower dynamics of the bleach signal (Fig. S4B).

The broadband near-infrared pump-probe measurements are a clear indication that the overall optical response of the sample to above-gap photo-excitations is similar, no matter how large the photon energy is. Only sub-gap photo-excitations trigger a suppression of the screening and unveil the exciton resonance in the monolayer.

## 5. Temperature dependence of the visible and near-infrared pump-probe signal

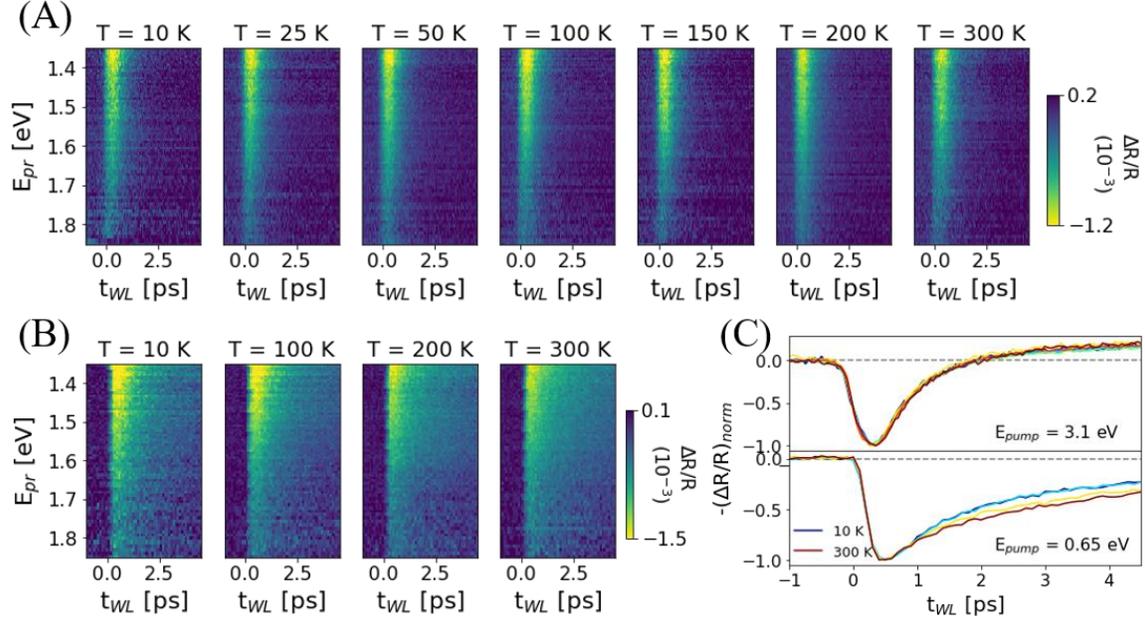

**Fig. S5: Transient reflectivity maps at different temperatures. (A), (B)** Transient change in reflectivity at different temperatures upon photo-excitation by visible ($E_{pump}$ = 3.1 eV) and near-infrared ($E_{pump}$ = 0.65 eV) pulses, respectively. **(C)** Normalized pump-probe traces at different temperatures integrated over the range 1.35-1.85 eV following a visible (top) and near-infrared (bottom) photo-excitation.

As highlighted in Supplementary note 1, BP features anomalous thermoelectric properties. In contrast to the vast majority of semiconductors, the gap energy in bulk BP monotonically increases with increasing temperature, as also observed in our sample through FTIR measurements (Fig.S1B). While different explanations have been proposed to describe such singular behaviour (i.e., gap-renormalization by strong electron-phonon coupling, thermal expansion [7], temperature-tunable interlayer coupling [10]), no unanimous consensus has been reached so far.

In the main text, we showed that the temperature-dependence of the electronic structure significantly affects the optical response to mid-infrared photo-excitation (Fig. 3). At fixed photon energy of the mid-infrared pulse, the photo-induced undressing of the excitonic resonance is hindered at higher temperature, where the gap energy becomes much larger than the photo-injected excess energy.

On the contrary, the optical response to visible and near-infrared pulses does not have such strong dependence on the sample temperature. We summarize in Fig. S5 the pump-probe measurements performed on bulk BP at different temperatures upon visible (Fig. S5A) and near-infrared (Fig. S5B) photo-excitation. The normalized visible pump-probe traces integrated over a broad energy range almost overlap at different temperatures (Fig. S5C, top panel). The same analysis on the reflectivity maps following a near-infrared photo-excitation (Fig. S5C, bottom panel) shows that also in this case the overall signal is similar at different temperatures. We observe a slower decay time of the photo-bleaching at higher temperatures, possibly due to a modified electron-phonon scattering rate.

## 6. Fluence-dependence of the visible pump-probe signal

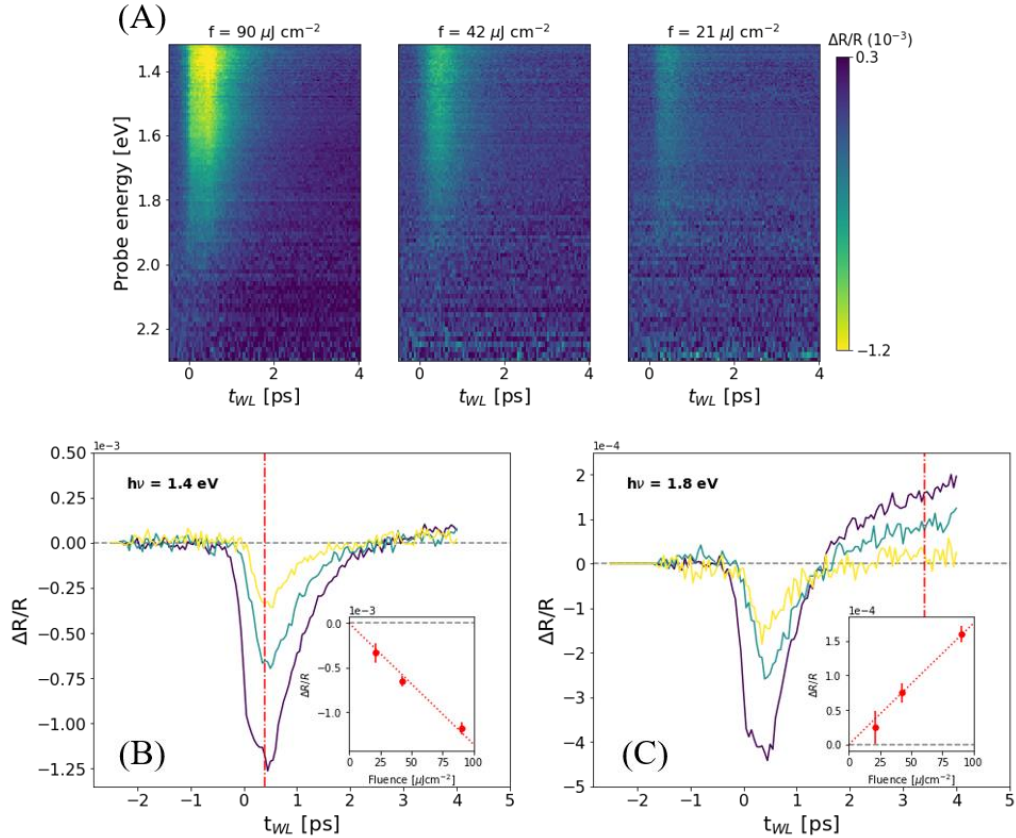

**Fig. S6: Fluence-dependent visible pump-probe measurements.** **(A)** Color-coded maps of transient change in broadband reflectivity after photo-excitation by visible ultrashort pulses with variable fluence at 10 K. **(B),(C)** Time-traces at selected probe photon energies ($h\nu$=1.4 eV and $h\nu$=1.8 eV, respectively) for all the three fluences under study. The insets show the fitted linear dependence of the transiently induced transparency at $t_{WL}$= 400 fs and $t_{WL}$= 3.5 ps.

We show in Fig. S6 the transient reflectivity upon photo-excitation by 3.1 eV pulses with different fluence. Both the photo-induced transparency by photo-bleaching (Fig.S6B) and the photo-induced absorption (Fig.S6C) scale linearly in pump fluence, as shown in the insets. This is an indication that we work in a regime in which the photo-excited free carrier population is proportional to the absorbed power.

## 7. Fluence-dependence of the mid-infrared pump-probe signal

We analyse in Fig. S7 the differential reflectivity following photo-excitation by MIR pulses with tunable fluence. The optical response at the lowest fluence under exam is the one discussed in the main text, where the only contribution to the signal is well localized in frequency and corresponds to the exciton resonance. When the MIR pump fluence is increased (up to almost an order of magnitude), a background signal arises on top of the exciton, whose spectral dependence (Fig. S7B) is similar to the one measured upon above-gap photo-excitation (Fig. 2B, yellow trace). We ascribe this frequency-broad background to non-linear two-photon absorption (TPA). In order to isolate the two contributions, we integrated the differential maps in Fig. S7A over two different energy regions: the exciton resonance lies in the range 1.45-1.9 eV, while the TPA signal dominates the low-energy side of the spectrum (1.3-1.45 eV). We plot in Fig. S7D,E the pump-probe traces integrated over these two spectral regions and normalized over the corresponding fluences. While the exciton resonance is characterized by a fast decay of approximately 1 ps, the TPA dynamics at high fluences features a slower decay time, which is very similar to the one measured upon near-infrared (0.65 eV) photo-excitation (Fig. S5C bottom panel). This reinforces our assignment of the TPA signal. A further confirmation comes from the fluence dependence of the exciton and TPA contributions (Fig. S7C). We fitted the points in the plot with a power law function ($\beta x^\alpha$) to extract the fluence dependence. While the TPA signal scales as the square of the fluence ($\alpha_{TPA} = 2.17 \pm 0.09$), in agreement with a two-photon process, the exciton resonance is consistent with a square root-like dependence ($\alpha_{exc} = 0.71 \pm 0.07$). This is an indication that the transient exciton absorption is the result of a coherent effect that scales with the amplitude of the electric field ($\sqrt{f} \sim |E|$). It should be noted that the two points at highest fluences of the TPA signal deviate from the expected quadratic power law. This behaviour can be explained by the saturation of optical absorption that has been observed in BP under strong illumination. Due to the Pauli blockade effect, interband transitions become forbidden and this results in a non-linear enhancement of transmittance in both few-layer and bulk BP [23].

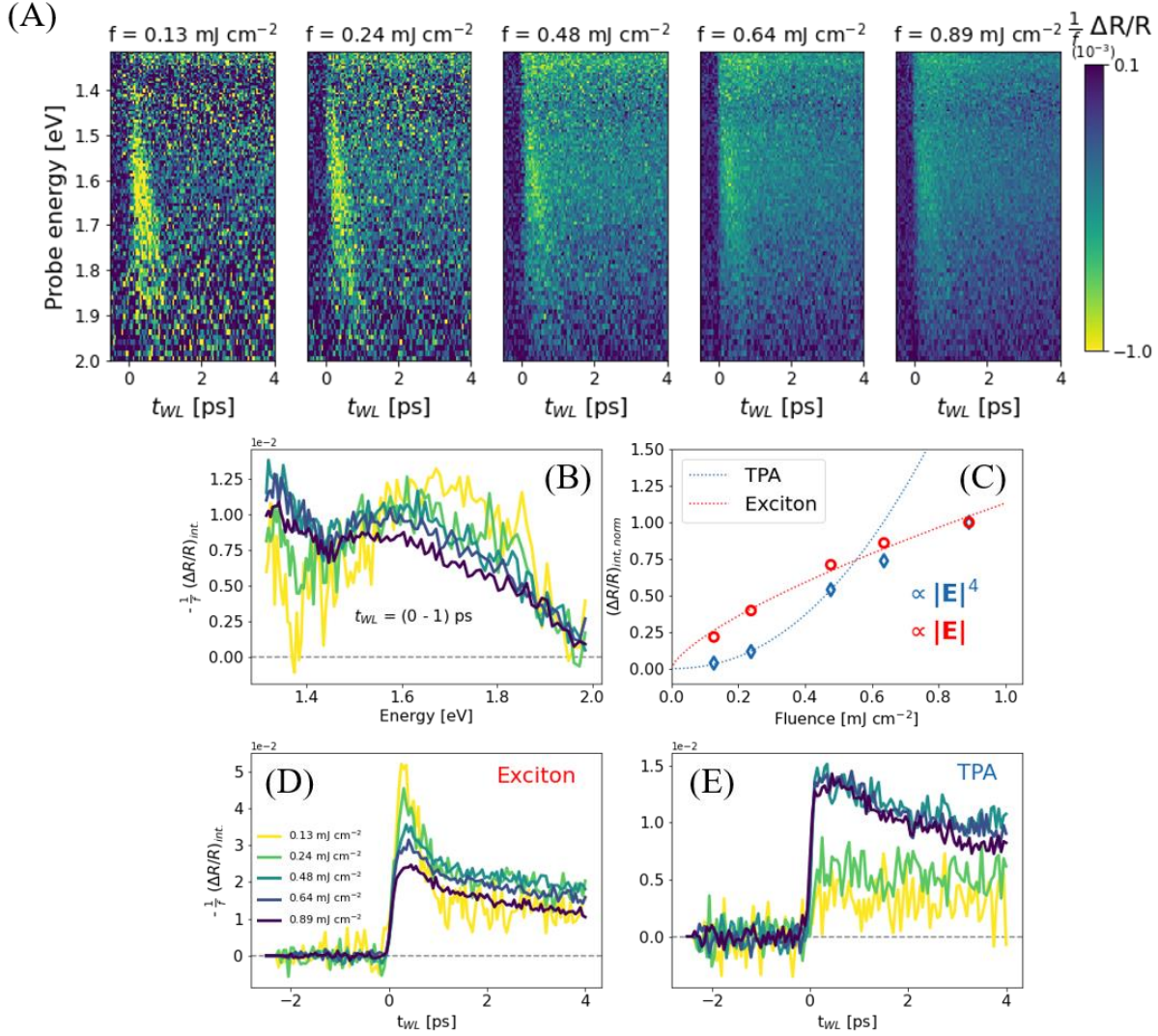

**Fig. S7: Fluence-dependent MIR pump-probe measurements.** (**A**) Transient reflectivity maps at 10 K upon photo-excitation by mid-infrared (275 meV) with different fluences ($f$). To highlight the exciton resonance at low fluences, all maps have been rescaled by the corresponding pump intensity ($\frac{1}{f}(\Delta R/R)$). (**B**) Energy spectra of the maps in (A) integrated over 1 ps after the photo-excitation and normalized over the pump fluence. (**C**) Normalized fluence dependence of the exciton and TPA contribution. Each point in the graph is obtained by integrating the spectra in (B) over the region 1.45-1.9 eV for the exciton, and 1.3-1.45 eV for the TPA signal. The dotted lines are power law fits to the data. We included in the TPA fit only the three measurements at lowest fluences. (**D**) Time-traces integrated over the exciton resonance (1.45-1.9 eV) normalized over the corresponding pump fluence. (**E**) Time-traces integrated over the energy region 1.3-1.45 eV to highlight the two-photon absorption (TPA) contribution.

## 8. Estimation of penetration depths, c-axis thermal expansion and local heating

We discuss in this section the calculation of some relevant optical quantities.

The penetration depth of the beams at a specific pump photon energy ($\delta_{PE}$) was calculated as:

$$\delta_{PE} = \frac{1}{\alpha} = -\frac{d}{\ln(T_{PE})}$$

(Eq. S2)

Where $\alpha$ is the absorption coefficient, $T_{PE}$ is the measured transmissivity at the photon energy considered, and $d$ is the thickness of the sample. Based on the measurements reported in ref. [24] and [25], we estimate the following penetration depths: $\delta_{3eV}$ =76 nm, $\delta_{0.65eV}$ =84 nm and $\delta_{275meV} = 1.5$ µm.

The interlayer spacing in bulk BP is 1.0473 nm, as measured through neutron diffraction [26]. We estimate the temperature jump required to increase the interlayer distance by 1% as:

$$\Delta T = \frac{\Delta L}{L_0}\frac{1}{\alpha_c} \simeq 850\ K$$

(Eq. S3)

Where $\frac{\Delta L}{L_0} = 0.01$ and $\alpha_c$ is the linear thermal expansion coefficient along the stacking direction measured in ref. [27].

Finally, we estimate the local heating induced by the mid-infrared pulse ($\Delta T_{MIR}$) as follows:

$$\Delta T_{MIR} = \frac{f(1-R)}{\delta_{275meV}\ C_{mol}}$$

(Eq. S4)

Where $f$ is the pump fluence, $R$ is the reflectivity at 275 meV (measured in ref. [28]), and $C_{mol}$ is the molar heat capacity (measured in ref. [29]). At the lowest mid-infrared pump fluence discussed in the main text ($f = 0.13\ mJ\ cm^{-2}$), we estimate a local temperature increase equal to about 7 K. We used in the calculation the BP density ($\rho_{BP} = 2.34$ g/cm³) and the BP exact molar mass (30.97 g/mol).

## 9. Reflectivity maps as function of MIR pump photon energy

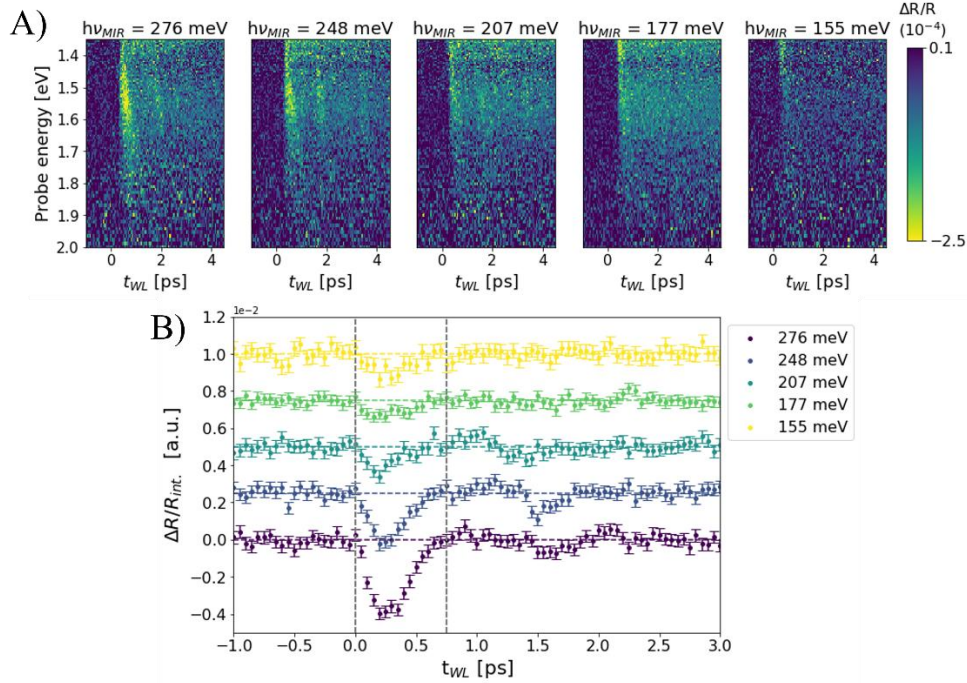

**Fig. S8: Photon energy-dependent MIR pump-probe measurements.** (**A**) Transient reflectivity maps upon photo-excitation by MIR pulses with tunable photon energy at constant fluence (160 μJ cm$^{-2}$). (**B**) Pump-probe traces integrated over the spectral region (1.45-1.65) eV after the subtraction of a slow-decay component through bi-exponential fitting of the data in (A).

We show in Fig. S8A the time- and energy-resolved transient reflectivity upon photo-excitation by MIR pulses with photon energy tunable across the bulk BP band gap. The measurement at the highest MIR photon energy displays a prominent contribution of the spectral feature peaked at 1.7 eV and assigned to the lowest-energy exciton resonance in the monolayer phosphorene. A spectrally-flat background is present, along with a replica of the signal at $t_{WL}$~2 ps that arises from a partial reflection of the copper substrate. As the MIR photon energy is decreased, the contribution of the exciton is reduced in intensity, but the flat background is not suppressed. In order to isolate the exciton contribution and quantify it as function of the MIR photon energy (Fig. 3A in the main text), we performed a bi-exponential fit of the pump-probe traces at each probe energy. We identified a fast-decaying component associated to the exciton resonance, and a slow-decaying component that is associated to the background and that has been subtracted. We show in Fig. S8B the energy-integrated pump-probe traces after the subtraction of the background. Fig. 3A is the result of the integration of these curves in the temporal window $t_{WL}$=(0-750) fs, as indicated in Fig. S8B by the dashed grey lines.

## 10. Mid-infrared pump-probe signal as function of the sample temperature

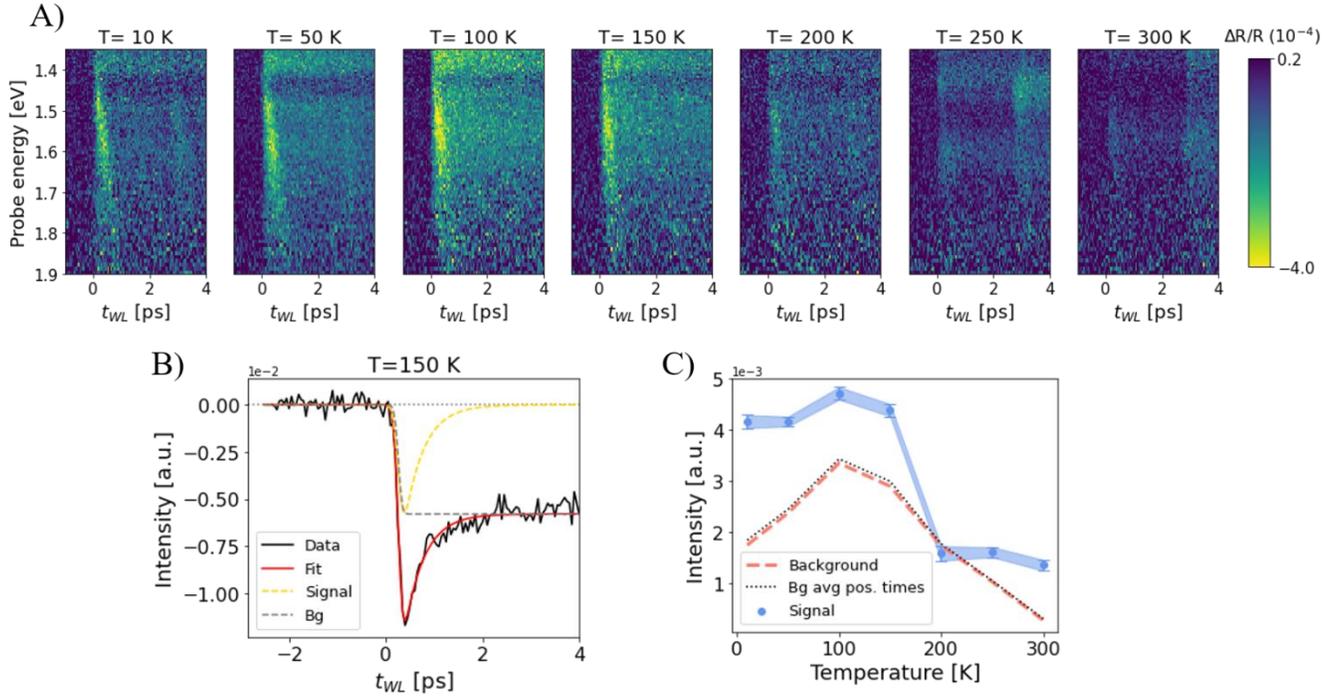

**Fig. S9: Temperature-dependent MIR pump-probe measurements** (**A**) Transient reflectivity maps upon MIR photo-excitation (hν=275 meV) at different temperatures. The fluence has been kept constant and equal to 160 µJ cm$^{-2}$. (**B**) Example at T=150 K of the analysis performed. The black curve is the pump-probe traces integrated over the spectral region 1.45-1.8 eV. The red curve is a fit to data that is the sum of a fast exponential decay (yellow curve) and a slow-decaying background component (error function, grey curve). (**C**) Fitted amplitude of the signal (light blue) and of the background (pink) as function of sample temperature. The dashed black curve indicates the fit-independent background obtained by integration of the pump-probe traces at positive times (3-4 ps).

We plot in Fig. S9A the time- and energy-resolved transient reflectivity maps upon photo-excitation by MIR pulses ($h\nu$=275 meV) at different sample temperatures. The $\Delta R/R$ map at low temperature consists of the frequency-localized signal associated to the phosphorene $E_{11}$ exciton resonance and a spectrally-flat background. Similar to the MIR photon-energy dependent measurements (Fig. S8A), the excitonic resonance disappears at higher temperatures, while the background persists, even if reduced in intensity. In order to study the temperature dependence of the two contributions independently, we analyzed the data as follows. Firstly, we performed an integration of the pump-probe traces over the spectral range of the exciton resonance (1.45-1.8 eV) and subtracted the spurious replica of the signal at $t_{WL} \simeq 3$ ps coming from the back-reflection of the copper substrate. The result is the black curve plotted in Fig. S9B for a representative map at T=150 K. We fitted this curve with the sum of a fast-decaying exponential (yellow curve), which reproduces well the dynamics of the exciton, and an error function (grey curve) to fit the background. By performing the same analysis on all the maps in Fig.S9A, we plotted in Fig.S9C the temperature-dependent amplitude of the excitonic signal (light blue curve) and of the background (pink curve). The latter is consistent with a fit-independent integration of the pump-probe traces at positive time delays (3-4 ps, dashed black curve). The fitted amplitude of the signal (light blue curve) is the temperature-dependence at $h\nu$=275 meV discussed in the main text (Fig. 3B).

### 11. Contributions to calculated optical absorption

In this section, we identify the electronic transitions responsible for the absorption spectrum (computed at the independent particle level) in the range [0-3.0] eV for bulk BP. Specifically, we consider various sub-regions of the absorption spectrum and determine the pairs (*ck,vk*) of a conduction and a valence state at a given *k*-point in the Brillouin zone, for which:

- The energy difference $E_c(k) - E_v(k)$ (transition energy) is within the energy range of the considered sub-interval of the absorption spectrum;
- The dipole matrix elements $|r_{vc}^x|$ (computed along the *x*-direction) are the most intense.

In this way, it is possible to identify which regions of the Brillouin zone contribute to a given portion of the absorption spectrum, selecting transitions characterized by a high dipole strength.

The left panels in Fig. S10 show the absorption spectrum computed for light polarized along the *x* direction: in each subfigure we highlight in red the portion of spectrum mainly determined by the transitions between valence and conduction states denoted by the arrows in the band structures shown in the corresponding panels on the right; the width of the arrows (denoting different transitions) is proportional to $|r_{vc}^x|$, normalized to the largest dipole matrix element in a given region. Further, the *k*-points, along which the band dispersions are shown, are expressed in reciprocal lattice units (rlu).

Fig. S10A,B show that the main contribution to absorption in the range [0.3,1.0] eV comes from transitions between the last occupied valence and the first unoccupied conduction bands, with both *k* parallel to the ΓY direction in the $k_z = 0.0$ plane and with *k* parallel to the ΓZ direction.

At higher photon energies (see Fig. S10C,D, Fig. S10E,F and Fig. S10G,H), the absorption is explained in terms of transitions between the same valence-conduction band pairs considered before, but characterized by wave vectors with progressively increasing $k_z$ (the in-plane component remains close to Γ and mostly parallel to ΓY direction).

Looking at Fig. S10I,L, we can rationalize the absorption structure at about 2.2 eV as a transition between the penultimate occupied valence and the second unoccupied conduction bands, in the $k_z = 0.5$ rlu plane of the Brillouin zone, where the last two valence bands (along with the first two conduction bands) are degenerate in energy.

Finally, Fig. S10M,N demonstrate that the main contributions to the absorption peak between 2.7 and 3.0 eV come from transitions between the penultimate valence and the second unoccupied conduction band, at the wave vectors at which the transitions responsible for the absorption in the [0.3,1.0] range occur.

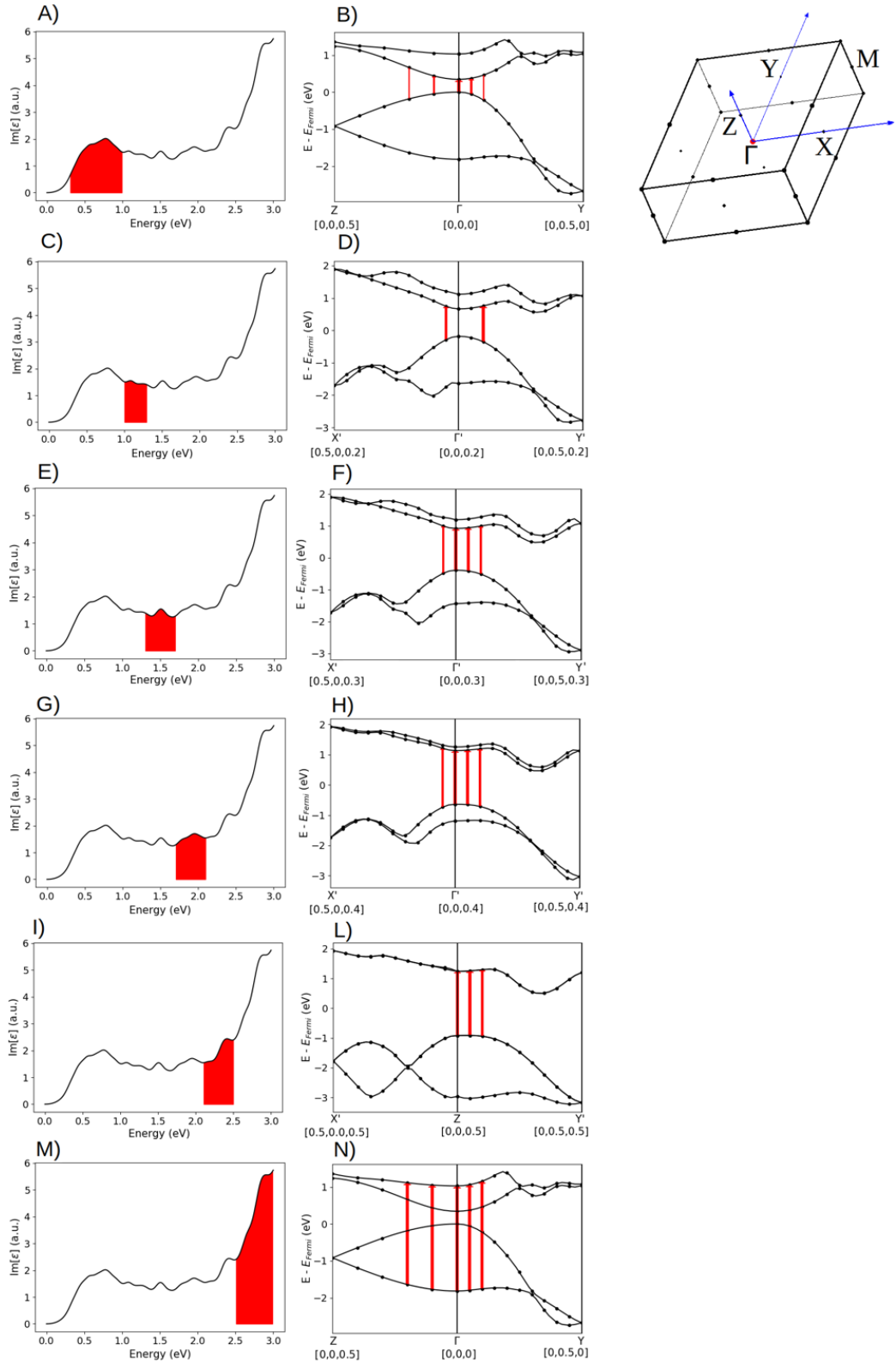

**Fig. S10:** Left panels show the absorption spectrum (i.e. the imaginary part of the dielectric function) for light polarized along the *x*-axis. In the right panels, we highlight the transitions along the high symmetry directions in the Brillouin zone that are responsible for the absorption in the red-shaded area of the spectrum in the corresponding left panels. See text for details. The high symmetry points are expressed in reciprocal lattice units.

## 12. Hydrodynamic pressure-dependent DFT calculations

We present in this section the effect of external hydrodynamic compression on the electronic properties of BP. More precisely, we focus on applied pressure values in the range between 0.2 GPa up to 1.0 Gpa. The lattice parameters in presence of an external hydrodynamic pressure have been determined by minimizing the Enthalpy of the structure. This has been done at the DFT level using PBE exchange correlation functional. The atomic positions within the unit cell have been relaxed following the procedure outlined in 'Methods' section. Finally, the electronic properties as a function of the external pressure have been computed with Hybrid DFT, using GAU-PBE hybrid functional.

In the left panels of Fig. S11 we summarize the effect of the application of an external pressure on the lattice parameters. The resulting in-plane contraction is anisotropic: it is more prominent along the armchair direction ($a$), while less important along the zigzag direction ($b$). As discussed in the main text, the electronic band gap is significantly reduced upon the hydrodynamic compression (right panel in Fig. S11).

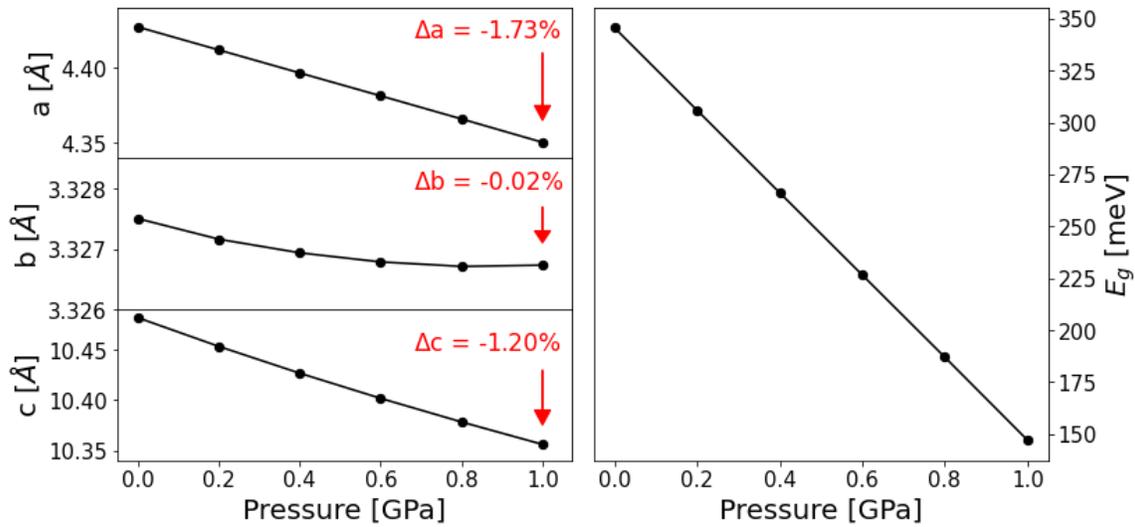

**Fig. S11:** The left panels show the compression of the lattice parameters as function of the applied external pressure. We highlighted the relative changes in the lattice parameters corresponding to the highest applied pressure, as denoted by the red arrows. On the right panel, the corresponding band gap energy is plotted as function of the pressure.

**Supplementary references**